\documentclass[aps,prl,twocolumn,superscriptaddress,nofootinbib]{revtex4}

\usepackage{epsfig}
\usepackage{graphicx}

\usepackage[centertags,sumlimits,intlimits,namelimits,reqno]{amsmath}
\usepackage{bbm,amsfonts}
\usepackage{amsthm}

\theoremstyle{definition}

\DeclareMathOperator{\tr}{tr}
\newcommand{\Ref}[1]{(\ref{#1})}

\def\N{{\mathbbm N}}

\def\>{\rangle}
\def\<{\langle}

\def\f{\frac}

\def\dag{^\dagger}
\def\what{\widehat}
\def\arr{\rightarrow}

\def\be{\begin{equation}}
\def\ee{\end{equation}}
\def\bes{\begin{eqnarray}}
\def\ees{\end{eqnarray}}

\newcommand{\no}{\nonumber\\}

\begin{document}
\title{Convergence Conditions for Random Quantum Circuits}
\author{Joseph Emerson\footnote{jemerson@perimeterinstitute.ca}}
\affiliation{Institute for Quantum Computing and Dept of Applied
Math, University of Waterloo}
\affiliation{Perimeter Institute for
Theoretical Physics}
\author{Etera
Livine\footnote{elivine@perimeterinstitute.ca}}
\affiliation{Perimeter Institute for Theoretical Physics}
\author{Seth Lloyd\footnote{slloyd@mit.edu}}
\affiliation{MIT, Dept of Mechanical Engineering}
\begin{abstract}
Efficient methods for generating pseudo-randomly distributed
unitary operators are needed for the practical application of Haar
distributed random operators in quantum communication and noise
estimation protocols. We develop a theoretical framework for
analyzing pseudo-random ensembles generated through a random
circuit composition. We prove that the measure over random
circuits converges exponentially (with increasing circuit length)
to the uniform (Haar) measure on the unitary group, though the
rate for uniform convergence must decrease exponentially with the
number of qubits.  We describe how the rate of convergence for
test functions associated with specific randomization tasks leads
to weaker convergence conditions which may allow efficient random
circuit constructions.
\end{abstract}

\date{\today}

\maketitle

Random unitary operators, consisting of unitary operators drawn
randomly from the invariant (Haar) measure on $U(D)$, comprise a
powerful resource for quantum computation and quantum
communication. Recent work has shown that random unitary operators
enable efficient protocols for characterizing noise (decoherence)
on universal quantum processing devices
\cite{Emerson03,EmersonQCMC,EAZ05}. Noise estimation is an
important problem for the experimental development of coherent
quantum control \cite{Pravia,Boulant} and, ultimately,
fault-tolerant quantum processing via the optimization of
error-correction schemes \cite{MikeandIke}. Moreover, random
unitary operators can be applied to randomize arbitrary quantum
state and consequently have proven useful for a number of quantum
communication protocols, including approximate quantum encryption
\cite{Hayden-epsRandom}, remote state preparation
\cite{Bennett-RSP}, and the superdense coding of quantum states
\cite{Harrow-Superdense}. A practical drawback for the above
applications is that a decomposition of the Haar-distributed
random operators in terms of one and two qubit gates (which is
required to generate a random unitary operator on a quantum
processor) requires quantum circuits of exponential length
$\mathcal{O}(D^2 \log(D)^3)$ \cite{Zyc2}. Such circuits are termed
\emph{inefficient} because the length grows exponentially with the
number of qubits ($\log_2(D)$).

In Ref.~\cite{Emerson03}, families of \emph{random circuits},
i.e., circuits that are composed of sequences of quantum gates
drawn independently and randomly from some universal gate set,
were proposed as a means of generating a set of
\emph{pseudo-random} unitary operators, i.e., a set that is
\emph{for certain practical purposes} indistinguishable from a
Haar-distributed set of random unitary operators. The possibility
that pseudo-random sets may be generated efficiently on a quantum
computer is suggested from previous work in quantum chaos, which
has shown that quantum chaos models reproduce relevant
characteristics of random unitary operators \cite{Emerson02} even
though these systems may be simulated with efficient quantum
circuits \cite{Shep01}. Moreover, numerical studies have shown
that random circuits of efficient length can reproduce that Haar
distribution of subsystem purity \cite{Emerson03,Cucchietti04},
and experimental evidence indicates that sufficiently random
operators can be generated with the currently available levels of
experimental control with NMR quantum processors
\cite{Emerson03,Ryan05}.

In this paper we develop a set of mathematical tools that can be
applied to analyze measures over random circuits. We apply these
tools to demonstrate that under increasing length of the random
circuit the measure over random circuits converges uniformly to
the Haar measure on the group. We show that the rate of
convergence (as a function of increasing gate depth) is
generically exponential, though the exponent must itself decrease
exponentially with the number of qubits. At the end of this paper
we describe how specific applications can define weaker
convergence conditions, leading to the possibility that efficient
random circuits composed from a discrete set of universal gates
might be able to adequately mimic the Haar measure.

We now prove that the measure of a product of unitary operators,
with each factor in the product drawn independently from a measure
enjoying support on a subset of $U(D)$ that generates the full
group (viz, a continuous universal gate set) or a dense subset of
the full group (viz, a discrete universal gate set), converges
exponentially to the constant function with respect to the Haar
measure on the group. Let $G$ denote the (compact Lie) group
$U(D)$ with elements $g \in G$. Let $\cal{F}$ denote the set of
probability measures over this group. Suppose the elements $g_1$
and $g_2$ are drawn at random from the measures $f_1 \in \cal{F}$
and $f_2 \in \cal{F}$ respectively. The composed element $g = g_1
\cdot g_2$, where $\cdot$ denotes the standard group
multiplication, is then distributed according to
\begin{equation}
f(g) = (f_1* f_2)(g)  = \int d\mu(h) f_1(g h^{-1}) f_2(h)
\nonumber,
\end{equation}
where $*$ denotes the convolution product and $d\mu(g)$ denotes
the Haar measure on $G$. The convolution operator for $m$
convolution factors is denoted $f^{*m} \equiv f*f^{*(m-1)}$.
Assuming that the initial function $f$ is a probability measure
then all its convolution powers $f^{*m}$ are also automatically
probability measures, and hence have the properties,
\bes
\forall g,\quad f^{*m}(g) & \geq & 0 \no \int_G d\mu(g)\,
f^{*m}(g)&=& 1. \ees We are interested in the convergence
properties of the sequence of probability distributions $f^{*m}$
as $m\arr \infty$. More precisely, we would like to prove its
convergence to the uniform function with respect to the Haar
measure and obtain an estimate of its rate of convergence. As an
aside let us first describe the intuition for expecting
convergence to the Haar measure. We note that only the characters
of the group are stable under convolution \cite{footnote}.
Therefore, if $f^{*m}$ converges at all then it must converge to a
character of the group. Moreover it must converge to the Haar
measure because this is the only character which is everywhere
non-negative. The hard part is now to determine under what
conditions the sequence $f^{*m}$ actually converges.

An important tool in the study of convolution products is the
Fourier decomposition. Indeed we will make use of the fact that
the convolution product turns into a simple standard product for
the Fourier components. In the case of compact Lie groups, a
Fourier representation of functions is provided by the Peter-Weyl
theorem (see e.g. \cite{Barut}). Let $\hat{G} = \{T^s \}$ be the
set of non-equivalent irreducible unitary representations of $G$.
We order the irreducible representations $s$ by increasing
dimensionality $d_s$, where the trivial representation has
dimensionality $d_{0} =1$. The functions $\sqrt{d_s} D_{jk}^s$
(taking $g \rightarrow \mathbb{C}$) for $s\in \hat{G}$ and $1 \leq
j,k \leq d_s$, where $D_{jk}^s(g)$ are the matrix elements of
$T^s$, form a complete orthonormal set for the Hilbert space
$L^2(G)$ of square-integrable functions on $G$ with respect to the
Haar measure. The orthogonality relations read:
\begin{equation}\label{orthonormality1}
\int_G d\mu(g)  D_{ij}^s(g) \overline{D_{mn}^{s'}(g)} =
\frac{\delta^{ss'}\delta_{im} \delta_{jn}}{d_s}
\end{equation}
where $\delta^{ss'}$ is 1 if $T^s$ and $T^{s'}$ are equivalent and
zero otherwise. Then, any function $f\in L^2(G)$ can be decomposed
as,
\begin{equation}\label{basis}
f(g) = \sum_{s\in \hat{G}} \sum_{j,k=1}^{d_s} \hat{f}_{jk}^s
D_{jk}^s(g),
\end{equation}
with Fourier coefficients
\begin{equation}\label{coeffs}
\hat{f}_{jk}^s = d_s \int_G d\mu(g) \; f(g) \;
\overline{D_{jk}^s(g)},
\end{equation}
and in particular, \bes \int_G d\mu(g) f(g) &=& \hat{f}^{s=0}.
\ees We have the following integral identity: \bes \int_G d\mu(g)
|f(g)|^2 &= & \sum_{s\in \hat{G}} \f{1}{d_s}\,
\tr\left((\hat{f}^s)\dag\hat{f}^s\right) < \infty
\label{L2norm} \ees
Finally, the normalized constant function (with respect to the
Haar measure) has Fourier coefficients: \be \label{HaarFT1}
\hat{\mu}^{s=0} = 1, \quad \hat{\mu}^{s \ge 1}_{jk} = 0. \ee

In general the Fourier transform of the convolution product has a
coefficient matrix for each irreducible representation of the
group given by the product of the coefficient matrices from the
same irreducible representation of the Fourier transforms of the
original convolved functions. That is, given two functions
$\phi,\psi$, the Fourier transform of their convolution product is
given as: \be (\what{\phi*\psi})^s\,=\,
\f{1}{d_s}\what{\phi}^s\,\what{\psi}^s. \ee Therefore the
convoluted powers of the initial probability distribution reads:
\begin{equation}
\label{mconv} (\widehat{f^{*m}})^{s} = d_s
\left(\frac{\what{f}^{s}}{d_s}\right)^m
\end{equation}
where $(\widehat{f^{*m}})^{s}$ denotes the $d_s \times d_s$ matrix
of coefficients appearing in the Fourier representation of
$f^{*m}$ and $(\hat{f}^{s})^m$ denotes the $m$'th power of the
$d_s \times d_s$ matrix of coefficients appearing in the Fourier
representation of the initial distribution $f$. The normalization
condition implies $(\hat{f^{*m}})^{s=0}=1$.

We assume that the initial measure $f$ is continuous and enjoys
support only on some (small) subset of $G$ that generates the full
group $G$. This subset might have very small measure compared to
the full group; one example is the continuous gate set consisting
of all single qubit rotations coupled with a CNOT gate between all
qubit pairs (later we relax this assumption and consider
convergence conditions for discrete and finite universal gate
sets). We have the following lemma: the matrix $\hat{f}^{s}$ has
norm strictly less than $d_s$ for $s> 0$. To prove this we
consider the usual vector norm, $ | x|_2 \equiv \sqrt{\langle x |
x \rangle}$, and observe that for all vectors $x$ in the $s$
representation, we have \cite{Hannan}:
\begin{eqnarray*}\label{proof2}
    | \hat{f}^{s} x |_2
    & = & | d_s \int_{G} d\mu(g) \; f(g) \;
\overline{D}^{(s)}(g) \; x |_2 \\
& \leq &  d_s \int_{G} d\mu(g) \left|
f(g) \; \overline{D}^{(s)}(g) \; x \right|_2 \\
& \leq & d_s
\int_{G} d\mu(g)\; f(g) \; \| \overline{D}^{(s)}(g) \| \;  | x |_2
= d_s |x|_2
\end{eqnarray*}
where we have introduced the norm, $  \| D\| \equiv \max_{x\neq 0}
|D x |_2/|x|_2 $, and used that
$ \| D\|   = 1 $
since $D(g)$ is unitary.
Equality holds if and only if $D(g) x = \xi(g) y $ is true for all
$g$ for which $f(g)> 0$, where $\xi(g) \in \mathbb{C}$. Since any
$g \in G$ can be generated by the support of $f$, $D(g) x = \xi(g)
y $ must hold also for all $g \in G$. This implies we have a 1-D
representation of $G$ embedded in the \emph{irreducible}
representation $s$. Hence the equality is not saturated unless $s$
is the trivial (identity) representation. It follows that, when
$s\ne 0$, i.e. for all $d_s
>1$, we have  $\forall x,\, | \hat{f}^s x| < d_s |x|$, hence
\begin{equation}
\label{normbound} \| \hat{f}^s \| \equiv \max_{x\neq 0}
\frac{|\hat{f}^s x|_2}{|x|_2} =\max_{|x|_2 =1}|\hat{f}^s x|_2
< d_s.
\end{equation}

Our proof is not concluded: this bound does not guarantee uniform
convergence to the uniform measure (given by Eqs.~(\ref{HaarFT1}))
because the Fourier representation for generic functions involves
an infinite sum, and hence the eigenvalues of very ``high
frequency" Fourier modes can {\it a priori} come arbitrarily close
to 1. Nevertheless, a first remark is that polynomial functions
$f$ (technically polynomials in the matrix elements of $g$) are
described by a finite sum in the Peter-Weyl decomposition. For
such a case, denoting by $S$ the ``frequency cutoff" from the
maximal representation label for the given polynomial, we obtain
the bound
$$
\|(\hat{f}^{*m})^s\|\le d_s \alpha^m, \qquad {\rm with} \quad
\alpha\equiv \max_{1\le s\le
S}\left(\f{\|\hat{f}^s\|}{d_s}\right)<1,
$$
and, consequently, a proof of the exponential convergence of the
convoluted powers towards the uniform measure.

For $f$ more generally any continuous function, one can repeat a
similar analysis which guarantees a uniform approximation using
only a finite Fourier sum. Indeed for $f$ continuous and any
$\epsilon > 0$ there exists a finite $N_\epsilon$ (independent of
$g$) such that, for all $g \in G$  \cite{Barut}, \be
\label{uniform}
 \left|f(g)  - \sum_{s=1}^{N_\epsilon}  \sum_{j,k=1}^{d_s} \hat{f}_{jk}^s
D_{jk}^s(g) \right| \leq \epsilon. \ee Let us call
$f_{N_\epsilon}$ the truncated function with the representation
cut-off $N_\epsilon$. Then for all $g\in G$ and $m\ge 2$, we use
the triangle inequality to write:
\begin{eqnarray*}
|f^{*m}(g)  - 1| & \leq & |f^{*m}(g)  - f^{*m}_{N_\epsilon}(g) | +
|f^{*m}_{N_\epsilon}(g) -1| \nonumber \\
& \le & |f^{*m}(g)  - f^{*m}_{N_\epsilon}(g) | \\
& + & \left|\sum_{s>1}^{N_\epsilon}  \sum_{j,k=1}^{d_s} d_s^{1-m}
 ((\hat{f}^s)^m)_{jk} D_{jk}^s(g)  \right|.
\end{eqnarray*}
It is straightforward to bound the first term using the fact that
$f_{N_\epsilon}*f_{N_\epsilon} = f*f_{N_\epsilon}$. For all $g \in
G$ we have,
\begin{eqnarray*}
|f^{*m}(g)  &-& f^{*m}_{N_\epsilon}(g) | =
|f*(f^{*(m-1)}-f^{*(m-1)}_{N_\epsilon})(g)| \\
& \le&
 \int d\mu(h)\, |f(gh^{-1})| |(f^{*(m-1)}-f^{*(m-1)}_{N_\epsilon})(h)| \\
&\le& \left\|f^{*(m-1)}-f^{*(m-1)}_{N_\epsilon}\right\|_\infty \\
& \le & \dots
 \le \left\|f-f_{N_\epsilon}\right\|_\infty\le\epsilon,
\end{eqnarray*}
using the fact that $f$ is a positive function with integral equal
to 1. We bound the second term with the help of the inequality:
$$
\left|\tr\left((\hat{f}^s)^m D^s(g)\right)\right| \le d_s
\|(\hat{f}^s)^m\| \le d_s \|\hat{f}^s\|^m,
$$
where we use that the vectors of the unitary matrix $D^s(g)$  have
norm one. It directly follows that:
$$
|f^{*m}_{N_\epsilon}(g) -1| \le
\alpha_{\epsilon}^m\sum_{s>1}^{N_\epsilon}d_s^2,
$$
where we have defined
$\alpha_\epsilon=\mathrm{max}_{1<s<N_\epsilon}(\|\hat{f}^s\|
/d_s)$. As we have shown above, for all non-trivial $s$,
$\|\hat{f}^s\| <d_s$ so that $\alpha_\epsilon<1$. Therefore there
exists a integer $M$, which can be chosen larger than
$N_\epsilon$, such that $|f^{*m}_{N_\epsilon}
-1|_\infty\le\epsilon$ for all  $m\ge M$. Then it is obvious that
for all $m\ge M$, we have bounded:
$$
\|f^{*m} - 1\|_\infty \le 2\epsilon.
$$
This concludes the proof that $f^{*m}$ converges uniformly to the
constant probability measure on the group for any continuous
probability measure. More generally, we claim that the
convergence is exponential for any $f \in L^2$. Because of Eq.~\Ref{L2norm} we know
that $\tr((\hat{f}^s)\dag\hat{f}^s)/d_s$ goes to 0 when $s$ goes
to $\infty$. Specifically,
$\|\hat{f}^s\|\le\sqrt{\tr((\hat{f}^s)\dag\hat{f}^s)}$, and
therefore we have $\|\hat{f}^s\|/\sqrt{d_s}\arr 0$,
so we don't need to worry
about the norm of $\hat{f}^s/d_s$ getting close to 1 when
$s\arr\infty$. More precisely, there exists $S_\delta \in\N$ such
that $\|\hat{f}^s\|/\sqrt{d_s}\le\delta < 1$ for all $s >
S_\delta$. Then for all $s$ we have the bound
\begin{eqnarray*}
\|f^{*m}-1\|_\infty & \le & \sum_{s>0} d_s^{-(m-2)}
\| \hat{f}^s \|^m \\
& \le & \alpha_\delta^m \sum_{0<s\le S_\delta} d_s^2
 +  \delta^m \sum_{s > S_\delta} d_s^{-(m/2-2)}
\end{eqnarray*}
where we have defined,
$$
\alpha_\delta\,\equiv\, \max_{s\le S_\delta}
\left(\f{\|\hat{f}^s\|}{d_s}\right) <1,
$$
As long as $m> 6$ the sum over $s$ in the second term converges.
Indeed, as explained in \cite{Vilenkin}, the irreducible
representations of $U(D)$ are usually labelled by a couple of
integers $(k,l)$ and their dimension is given in terms of binomial
coefficients:
$$
d^{(D)}_{k,l}=\f{k+l+D-1}{D-1}C^k_{k+D-2}C^l_{l+D-2}.
$$
It is then straightforward to check that the dimensions grow
sufficiently rapidly with $s$ to make the sum over $s$ converge.
The exponential $\alpha^m$ defines the convergence rate of the
convoluted powers of $f$ to the uniform probability measure.


We have shown that the measure over random circuits converges to
the Haar measure and moreover that the rate of convergence (with
increasing circuit length $m$) is exponential. However the
exponent will generally depend on the Hilbert space dimension $D$.
It is clear that for uniform convergence the exponent must
decrease at least as rapidly as $D^2 \log(D)^3$. This follows from
the fact that $\mathcal{O}(D^2 \log(D)^3)$ gates are required to
generate all the elements of $U(D)$ from a fixed universal gate
set \cite{MikeandIke}. However, the requirement of \emph{uniform}
convergence we have considered is much stronger than necessary for
any practical application. Indeed the practical statistical
distinguishability will be limited by practical constraints, such
as bounded computational resources or finite sampling from the
distribution. In particular we are concerned with convergence with
respect to some operationally restricted class of test functions.
Hence we demand only that $f^{*m}$ converge to the Haar measure
for the "weak topology", i.e. $\int_G\,d\mu(g)\,f^{*m}(g) \phi(g)$
converges to $\int_G\,d\mu(g) \,\phi(g)$ for appropriate
(continuous) test functions $\phi$. A first example of an
operationally motivated test function is the fidelity loss under
the motion reversal of a random unitary \cite{EAZ05}. Another
example comes from the important case of the distinguishability of
quantum states evolved under unitary operators drawn from
different distributions. Both criteria lead to test functions
$\phi$ which are given by polynomials in the matrix elements of
the fundamental irreducible (unitary) representation $D^{s=1}(g)$
of $g$, and the finite degree of the polynomial fixes a Fourier
cutoff. Hence the convergence is automatically an exponential and
the rate of convergence may lead to efficient (scalable) random
circuit constructions. Convergence for the weak topology (i.e.,
for polynomial test functions) also allows us to extend our
analysis to arbitrary distributions $f \in L^1$, such as finite
sums of $\delta$ functions.
The same technique applies and we only need to require that
$f$ has support on a discrete universal gate set.
The argument leading to Eq.~\ref{normbound} still holds and
exponential convergence follows for the weak topology, as explained above.

In general then the rate of convergence will depend on the
convergence condition specified from an appropriate test function
and on the initial probability distribution $f$. The initial
distribution $f$ can be modelled in a variety of ways. For
example, if we start with a Gaussian packet, then its width will
simply increase linearly with the power of the convolution,
eventually converging to the uniform measure. The exact
convergence rate is then determined directly from the maximum
matrix norm of the Fourier components of the initial distribution,
where the maximum is taken only over those Fourier components
below the cutoff determined from the test function. The
appropriate test function (and Fourier cutoff) will generally
depend on the specific application. Indeed the numerical analysis
of Refs.~\cite{Emerson03,Cucchietti04} indicates that for a test
function given by the subsystem purity the rate of the exponential
convergence is asymptotically independent of the dimension $D$,
suggesting that random circuits can generate Haar-distributed
subsystem purity efficiently.

We would like to acknowledge R. Cleve, D. Cory, D. Gottesman, J.
Goldstone, A. Harrow, H. Pfeiffer, and P. Zanardi for helpful
discussions.


\begin{thebibliography}{99}
\bibitem{footnote} More precisely, if we label the irreducible
representations of $G$ by $s$, then the functions
$g\mapsto\tr_s(gP)$, with $P$ an arbitrary projection on the
Hilbert space of the $s$ representation, or any linear combination
of such functions (for distinct $s$'s), are the only functions
invariant under the convolution product. The characters are the
special case when the projection is trivial $P={\rm Id}_s$.
Nevertheless, it is still clear that none of these functions are
non-negative except for the trivial representation $s=0$.
\end{thebibliography}
\end{document}